\documentclass[longbibliography,prl,aps,twocolumn,showpacs]{revtex4-1}
\usepackage{amsmath}
\usepackage{amssymb}

\usepackage[pdftex,dvipsnames]{xcolor}
\usepackage{xargs}
\usepackage[colorinlistoftodos,prependcaption,textsize=tiny]{todonotes}

\usepackage{lipsum}
\usepackage{graphicx}
\graphicspath{{figs/}}

\begin{document}

\title{Velocity distribution of a homogeneously driven two-dimensional granular gas}

\author{Christian Scholz}
\affiliation{Institute for Multiscale Simulation,  Friedrich-Alexander-Universit\"at Erlangen-N\"urnberg, Germany}

\author{Thorsten P\"oschel}
\affiliation{Institute for Multiscale Simulation,  Friedrich-Alexander-Universit\"at Erlangen-N\"urnberg, Germany}

\date{\today}

\begin{abstract}
The theory of homogeneously driven granular gases of hard particles predicts that the stationary state is characterized by a velocity distribution function with overpopulated high-energy tails as compared to the exponential decay valid for molecular gases. While this fundamental theoretical result was confirmed by numerous numerical simulations, an experimental confirmation is still missing. Using self-rotating active granular particles, we find a power-law decay of the velocity distribution whose exponent agrees well with the theoretic prediction.
\end{abstract}
\maketitle 

{\em Introduction.} 
Granular gases establish the dilute limit of granular matter. While in many granular systems the particles interact via long-lasting or permanent contact, in granular gases the particles follow ballistic trajectories, interrupted by pairwise dissipative collisions. In a model description, the particles are ideally hard, 
such that collisions are instantaneous events. Apart from being rather successful to explain the physics of {\em rapid granular flows} \cite{Goldhirsch:2003a} (and many references therein), the concept of granular gases is also of practical importance for the description of natural phenomena, such as the rings of the large planets of our Solar system, e.g. \cite{Spahn:2006}.  Granular gases have been intensively studied for their rich phenomenology, most prominent the cluster and vortex  instabilities \cite{Goldhirsch:1993a,Noije:1997}, which do not have corresponding equivalents in the physics of molecular gases. 

{\color{black}To date} there is no canonic theory for granular gases, 
however, much progress was achieved for the {\em homogeneous force-free case}, that is, infinitely extended systems in the absence of external forces and boundaries due to confinement. Here the mathematical tools known from the {\color{black}kinetic theory} of molecular gases, in particular the (pseudo-) Liouville-operator and the Boltzmann-Enskog equation may be adopted \cite{Brey:1997,Goldhirsch:2003a,Noije:1997b,SMGG} to granular gases which are {\em always} in non-equilibrium. Based on this approach, 
the full hydrodynamics of granular gases can be derived, \cite{Sela:1998,Brey:1998,Goldhirsch:2003a,Brilliantov:2003}. 
{\color{black}Although} the homogeneous force-free granular gases are phenomenologically similar to molecular gases, there are numerous phenomena arising from the dissipative nature of particle interaction such as anomalous diffusion \cite{Brilliantov:2000a,Brilliantov:2005}, correlations between the translational and rotational degrees of freedom \cite{Brilliantov:2007a,Kranz:2009} and others.

The most fundamental characteristics of a granular gas is the velocity distribution function.
{\color{black}For $v\sim v_T$, where $v_{T} = \sqrt{\left<v^2\right>/2}$  is the thermal velocity, it is slightly distorted with respect to the Maxwell distribution which may be described by a low-order Sonine expansion \cite{Goldshtein:1995}}. For large velocities, $v\gg v_T$, however, the distribution differs qualitatively from the Maxwellian, $f(v)\sim\exp(-v^2)$, by revealing an overpopulated high-energy tail, $f(v)\sim \exp(-v)$ \cite{Esipov:1997}. This tail leaves a pronounced fingerprint in the hydrodynamic fields \cite{Poeschel:2006,Poeschel:2007}. By now, the overpopulated tail of the distribution function is theoretically well accepted as it was reproduced using different theoretical approaches \cite{Goldhirsch:2003b,Noskowicz:2007,Villani:2006} and numerical simulations, e.g., \cite{Esipov:1997,Brey:1999,Puglisi:1999,Moon:2003,Brey:1999a,vanZorn:2005}.

The experimental investigation of the homogeneous force-free state is difficult for several reasons: (a) starting from some initial velocity distribution, e.g. Maxwellian, the gas needs time  (measured in collisions per particle) to develop its native distribution \cite{Poeschel:2007}. At the same time, the thermal velocity reduces exponentially {\color{black}due} to dissipative collisions. For realistic material properties, particle sizes, and collision velocities, there is only a small time window where the gas is still homogeneous and at the same time reveals a well developed distribution function \cite{Poeschel:2007}; (b) unlike numerical simulations where periodic boundary conditions may be modeled, experiments require always some confinement leading to inhomogeneous density. For instance, solid walls imply an enhanced collision rate and, thus, faster cooling and increased density close to the walls, e.g. \cite{Meerson:2004}. Therefore, experimental systems must be {\em very} large to have invariant conditions at least in the center region of the system, far from the confinement; (c) gravitation impedes force-free conditions. While (c) can be overcome by performing experiments in microgravity \cite{Grasselli:2015,Tatsumi:2009,Hou:2008}, arguments (a) and (b) persist.

A way out is to consider {\em homogeneously driven} systems. Here, random forces are applied to the particles, thus enforcing a stationary state. By the {\em same} mathematical analysis as for the force-free case, it was shown that the velocity distribution reveals the same properties \cite{van1998velocity}, in particular a high-energy tail, $f(v)\sim\exp(-v^\beta)$ with $\beta=3/2$. Therefore, a corresponding experiment showing the high-energy tail for a randomly driven system would support the theory of granular gases considerably.

Experiments in this direction have been done before \cite{Schmick:2008,Grasselli:2015,Tatsumi:2009,Rouyer:2000,Baxter:2003,Olafsen:1999,Losert:1999,Kudrolli:2000}, in particular measurements for sub-monolayers of spheres on vibrating tables. However, the application of random forces individually to the particles and independent of their position could not be achieved. Instead, in such systems, horizontal particle motion results from out-of-plane collisions of the particles and/or microscopic surface roughness, leading to undesired correlations of the particles. We believe that this is not a well defined way of creating a true two-dimensional granular gas with random collisions, thus, the true origin of overpopulated high-energy tails is difficult to pinpoint.


The groundbreaking idea came from Cafiero et al. \cite{cafiero2002} who supplied energy to the system by randomly exciting the {\em rotational} degree of freedom in a two-dimensional {\color{black}molcular dynamics} simulation {\color{black} and showed that the result is independent of the particular method by which energy is injected into the system}. The energy was then transferred to the linear velocities through thermalization between the rotational and translational degrees of freedom, thus no artificial correlations were introduced.

In this Letter, we describe an experiment where we investigate a driven two dimensional granular gas where the energy is supplied through random excitations of the rotational degree of freedom. Our result shows clear evidence of an overpopulated high-energy tail, thus confirming the results of {\color{black}kinetic theory} of granular gases in the homogeneous  regime.

\begin{figure}[tbp]
\centering
  \includegraphics[width=.98\columnwidth]{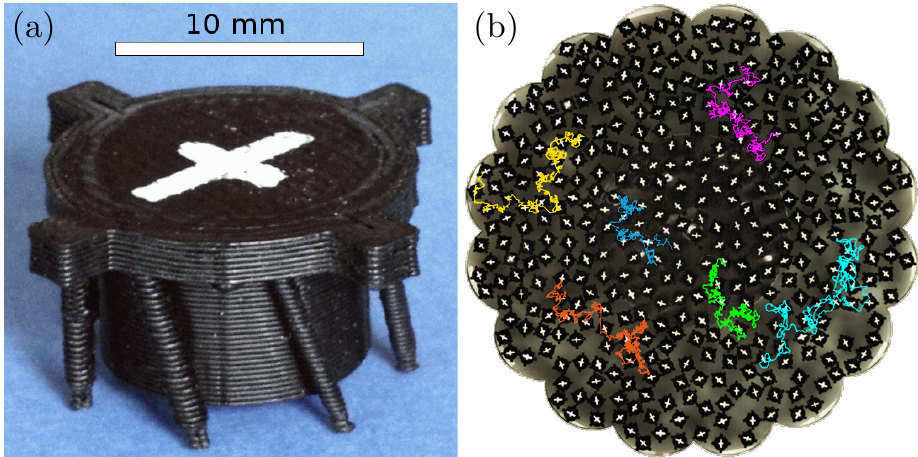}
  \caption{(Color online) (a) An active particle (Vibrot) with seven legs manufactured by rapid prototyping. A disk of diameter 15\,mm and width 2\,mm rests on 7 legs of length 8.5\,mm tilted by $18^\circ$. As second cylinder (height $6\,\text{mm}$, diameter $11\,\text{mm}$) is attached to the disk to lower the center of mass and stabilize the particle's motion. {\color{black}Four small cantilevers (3 mm) are attached to the edge.}  (b) Top view of the system with $N=379$ particles (packing fraction $\varphi=0.6$). The traces of some particles are overlaid.}
  \label{fgr:setup}
\end{figure}

{\em Experimental Setup.} 
The particles used in our two-dimensional experiment are disks resting on circularly aligned titled elastic legs (Fig. \ref{fgr:setup}a). 
When placed on an vibrating horizontal table, through an interplay of friction, inertial forces, and inelastic interaction between the particle and the vibrating wall \cite{Scholz:2016}, these particles suffer random excitations which puts them into irregular rotation. This type of particles was introduced by Altshuler at al. \cite{altshuler2013} and {\color{black}named} {\em Vibrot} since it transfers vibrational energy into rotation but not into linear motion in the plane of the table, {\color{black}due} to the symmetry of the particles. The intensity of the stochastic kicks can be controlled by the parameters of the vibration \cite{altshuler2013,Scholz2015}. It depends also on the geometry of the particle, in particular the number and inclination of the legs, and on the particle material \cite{Scholz:2016}. The particle shown in Fig.\,\ref{fgr:setup}a is decorated by a white cross on the top to conveniently determine its velocity and position and also by small cantilevers to enforce stochastic kicks and thus accelerate the thermalization between rotational and translational motion. Large numbers ($N\ge 500$) of such particles of high precision can be conveniently manufactured using 3d printers from acrylonitrile butadiene styrene (ABS) (density $\rho=1.07\,\text{g}/\text{mm}^3$, Young modulus $Y=2.3\,\text{GPa}$). The typical coefficient of restitution of the disks is $\varepsilon=0.5$ \cite{Scholz:2016}. 

In the experiment, the particles move on a smooth circular acrylic baseplate (diameter $480\,\text{mm}$, width $30\,\text{mm}$) attached to an electromagnetic shaker and leveled  horizontally to an accuracy of ${10^{-3}}^\circ$, to avoid gravitational drift. We used a flower shaped lateral confinement as {\color{black}used} by \cite{deseigne2010,deseigne2012vibrated} to avoid 
creeping motion along the container walls{\color{black}(see also \cite{tsai2005})} as the flower-like border would re-inject creeping particles back into bulk.
Figure \ref{fgr:setup}b shows a top view of the system. The traces of some particles are overlaid to show that the particles indeed perform a diffusive random walk driven by random kicks. No directed linear motion can be observed, see also  \cite{altshuler2013}.

The motion of the particles is recorded using a high-speed camera (500 frames/s). 
Using standard image recognition methods we track the positions of the particles with sub-pixel accuracy \cite{crocker1996methods} and subsequently we compute their linear and angular velocities.

Note that the system considered here is a two-dimensional granular gas. The vibration in vertical direction provides only the energy to drive the particles. Their motion in the horizontal plane is not directly affected by the vibration. Thus, the values of the amplitude and frequency are unimportant as long as they are chosen large enough such that the rotation of the particles is driven sufficiently and small enough such that the particles do not {\color{black}noticeably leave the horizontal plane}. The choice $f=50\,\text{Hz}$ and $A=0.17\,\text{mm}$ fulfills these conditions, that is, the results do not noticeably depend on $f$ and $A$ \cite{altshuler2013,Scholz2015,Scholz:2016}.  The same applies to the coefficient of restitution since the predicted overpopulated tail of the distribution function is invariant for any value of $\varepsilon$, e.g. \cite{Poeschel:2007}.

{\em One-particle properties.}
Before studying the velocity distribution function, we check carefully whether the preconditions for the comparison with the theoretical results are fulfilled, namely (a) single particles are required to obey a Maxwellian velocity distribution and (b) the system shall be homogeneous and isotropic.

\begin{figure}[htbp]
	\includegraphics[width=\columnwidth]{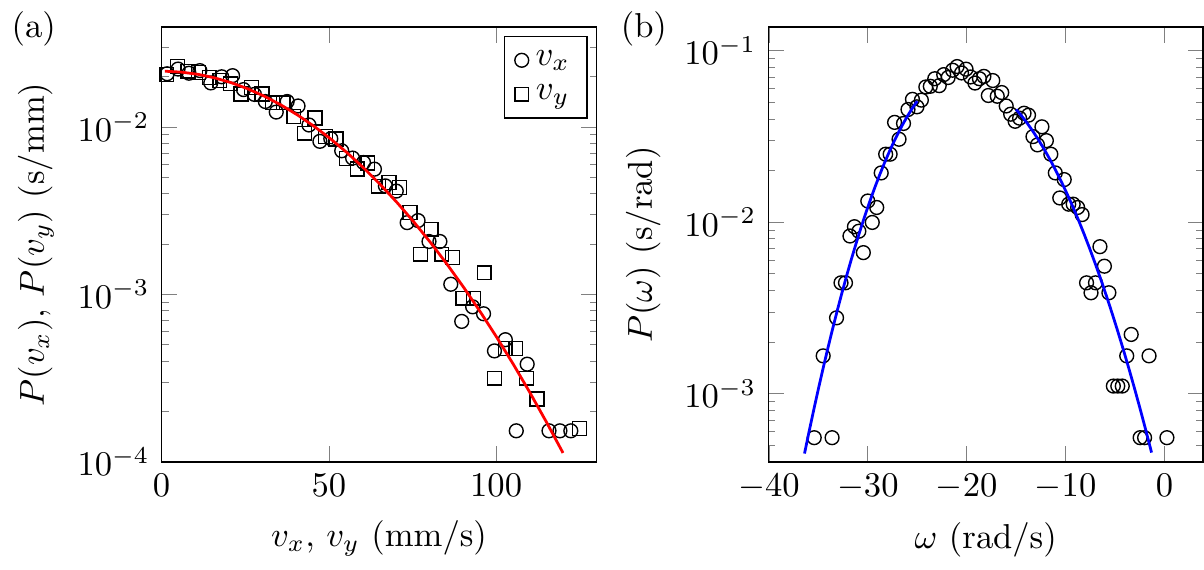}
	\caption{\label{Fig:SingleParticle} {\color{black}(a) Distribution of the absolute velocity components $v_x = |\vec{v} \cdot \vec{e}_x|$ and $v_y = |\vec{v} \cdot \vec{e}_y|$ of a single particle without interaction with other particles.} The solid line shows a fit to a Gaussian. b) Distribution of the rotational velocity {\color{black} of asymmetric Gaussian shape}  (solid lines).}
\end{figure}

Figure \ref{Fig:SingleParticle}a shows the velocity distribution of an isolated particle without interaction with other particles or the system boundary. The distribution of the $x-$component of the velocities (where $x$ is an arbitrarily chosen horizontal direction) is a Gaussian with very good accuracy (solid line in Fig. \ref{Fig:SingleParticle}a). Together with the isotropy of the system, to be shown below, this gives rise to a Maxwell distribution of the absolute values of the velocities.

For completeness, Fig. \ref{Fig:SingleParticle}b shows the distribution of the rotational velocity, $P(\omega)$, which is {\color{black}shifted} due to the asymmetric inclination of the particle's legs. Its tails follow a Gaussian. Note that $P(\omega)$ is irrelevant for the results presented here, due to the arguments given above. The important point is that the rotational driving leads to a Maxwell distribution of the linear absolute velocities, that is, the function $P(\omega)$ must not lead to deviations of the $P(v_x)$ from a Gaussian which is provided as shown in Fig. \ref{Fig:SingleParticle}a.

\begin{figure}[htbp]
\centering
  \includegraphics[width=\columnwidth]{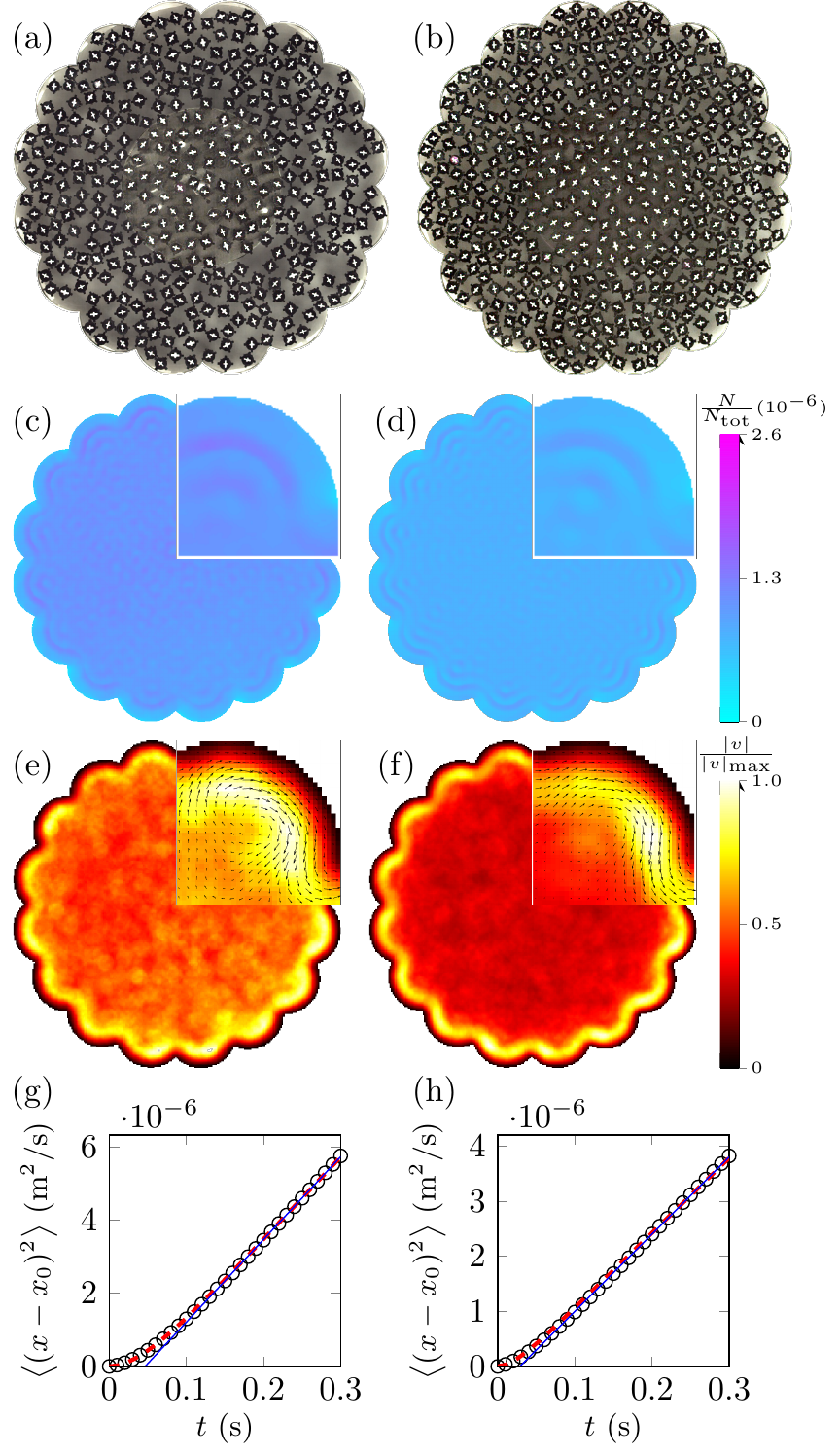}
  \caption{(Color online) (a,b) Snapshots (top-view) of the system at volume fractions $\varphi=0.47$ (top) and $\varphi=0.6$ (bottom), with the corresponding coarse grained fields of density (c,d) and velocity (e,f). The inset in (c-f) show a magnifications of a region close to the system border. The fields of density and velocity are homogeneous and isotropic, except for the wall region due to creeping particles \cite{deseigne2010,deseigne2012vibrated}. {\color{black}(g,h) Ensemble averaged mean squared displacement. For short times we observe the ballistic part of the motion (dashed curve), for longer times the particle motion becomes diffusive (solid line).}}
  \label{fgr:properties}
\end{figure}

{\em Homogeneity and isotropy.}
The collective behavior of the particles is studied for systems at two different packing fractions, $\varphi=0.47$ ($N=379$) and $\varphi=0.6$ ($N=479$). Snapshots of both systems are shown in Fig. \ref{fgr:properties}a,b.

By measuring the particle positions over an interval of 50 s and coarse graining {\color{black}using a Gaussian filter with width $R/2$, where R is the inner particle radius}, we obtain the normalized fields of density and absolute particle velocity, shown in Fig. \ref{fgr:properties}c-f. Both fields are isotropic and homogeneous, except for the region close to the system boundary. Here, the particles creep along the walls (in particular at convex walls) which can be seen from the vector arrows in the magnifications of Figs. \ref{fgr:properties}e,f. Similar effects have been observed for translationally active particles \cite{Kudrolli:2008,Kudrolli:2010,deseigne2010}. Due to our flower-like system walls, this effect is limited to about two layers of particles. {\color{black}The resulting collective rotation is negligible, since the average angular velocity of the entire system is four orders of magnitude below the individual spin of a particle.} Otherwise, the system is homogeneous down to less than $4\%$ {\color{black}(standard deviation of the velocity magnitude)}.\\
{\color{black} To ensure that the ballistic motion of the particles is captured, we show the ensemble averaged mean squared displacement in $x$-direction in Fig.~\ref{fgr:properties}(g,h). As desired, for short times, we observe the ballistic part (dashed curve) of the mean squared displacement and for longer times the diffusive motion (solid line). From the transition from ballistic to diffusive motion we obtain the relaxation time $\tau=47\,\text{ms}$ for $\varphi=0.47$ and $\tau=22\,\text{ms}$ for $\varphi=0.6$, which is one order of magnitude larger than our experimental temporal resolution. Furthermore, $\tau$ is significantly larger than the duration of a collision \cite{Schwager:2008},
  \begin{equation}
    \label{eq:4}
    \tau_c \approx 3.218\, R\frac{\left[ \sqrt{2}\pi\rho\left(1-\nu^2\right)\right]^{2/5}}{ Y^{2/5} v^{1/5}}\approx 15.4 \mu\text{s}
  \end{equation}
with the material parameter given above and $v=v_T$, such that non-binary collisions can be neglected.}

\begin{figure}[htbp]
\centering
  \includegraphics[width=\columnwidth]{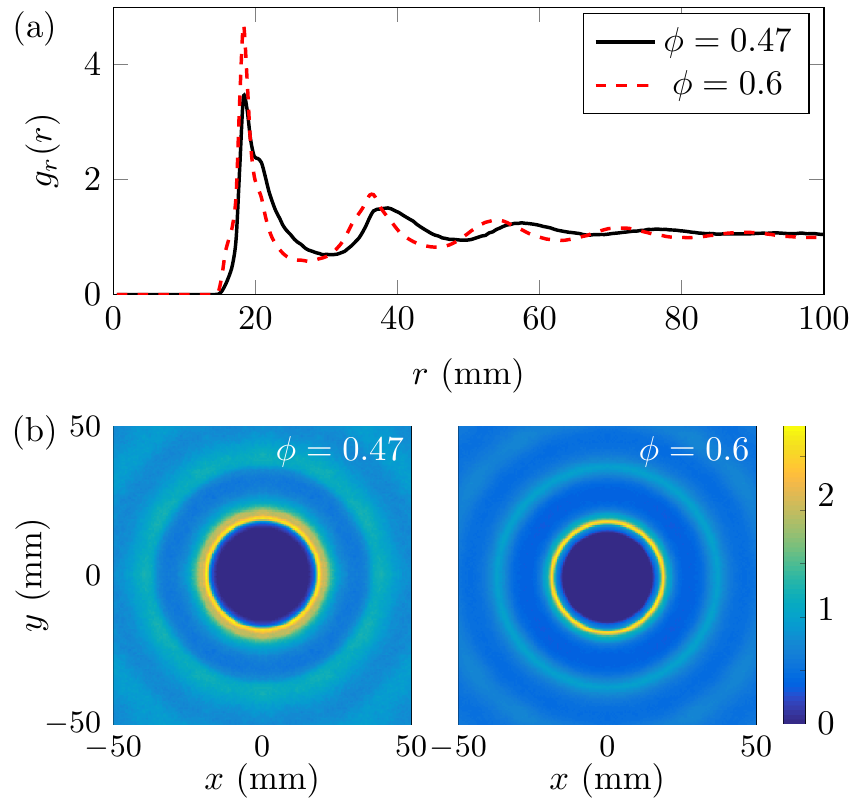}
  \caption{(Color online) (a) Radial pair correlation function $g\left(\left|\vec{r}_1-\vec{r}_2\right|\right)$ for filling fractions $\varphi=0.47$ and $\varphi=0.6$. (b) Vectorial pair correlation function, $g\left(\vec{r}_1-\vec{r}_2\right)$. No sign of clustering (long range correlations) nor anisotropy beyond statistical fluctuations can be seen.}
  \label{fig:gofr}
\end{figure}

Characterizing the spacial structure of the system by the (absolute) pair correlation function $g\left(\left|\vec{r}_1-\vec{r}_2\right|\right)$, (Fig. \ref{fig:gofr}a), we observe a fluid-like behavior, without long-range correlations. In particular, we do not see any sign of crystalline structure as it is typical for granular systems when undergoing clustering.  Further, no sign of anisotropy can be seen from the vectorial distribution functions, $g\left(\vec{r}_1-\vec{r}_2\right)$ shown in Fig. \ref{fig:gofr}b.

{\em Velocity distribution function.} After having shown in the last two sections that the experimental system indeed fulfills the preconditions regarding homogeneity, isotropy and single-particle behavior, for which the {\color{black}kinetic theory} of granular gases in the homogeneous state was derived, we consider now the velocity distribution function.

\begin{figure}[htbp]
\centering
  \includegraphics[width=\columnwidth]{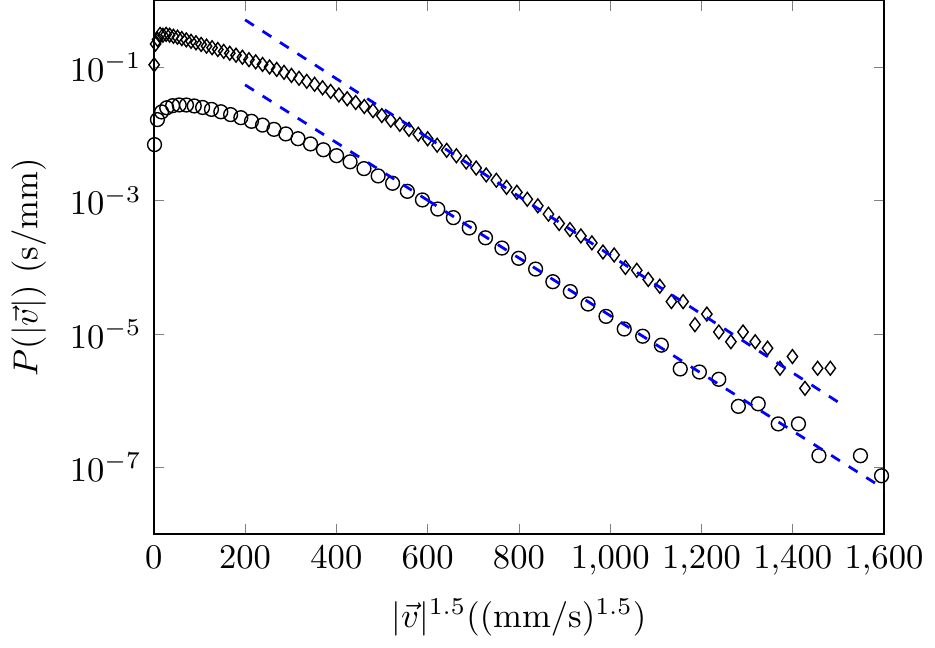}
  \caption{(Color online) {\color{black}Velocity distribution function $P\left(|\vec{v}|\right)$} for density $\varphi=0.47$ and $\varphi=0.6$. The dashed lines show the function predicted by {\color{black}kinetic theory} of granular gases \cite{Noije:1997}, $P(v) \sim e^{-v^{3/2}}$. 
  For better visibility, the curve for  $\varphi=0.47$ was shifted vertically by a factor of 10. {\color{black} Note that the shown velocity interval $0\le \left|\vec{v}\right| \lesssim 137\,\text{m/s}$ agrees with the interval of the one-particle velocity distribution shown in Fig. \ref{Fig:SingleParticle}a.}}
  \label{fig:distribution}
\end{figure}

Figure \ref{fig:distribution} shows the velocity distribution function,  $P\left(v\right)$, {\color{black}where $v=|\vec{v}|=\sqrt{{v_x}^2+{v_y}^2}$,} together with the predictions of the {\color{black}kinetic theory} of granular gases \cite{Noije:1997} for the high-energy tail, $P(v) \sim e^{-v^{3/2}}$.
The data shown in Fig.~\ref{fig:distribution} is based on the high-speed recordings of the system over a period of 50 s.

Clearly, the distribution functions are well compatible with the theoretical result, $P\left(v\right)\sim \exp\left(-v^\beta\right)$ with $\beta=3/2$ but not with  the equilibrium distribution of a molecular gas, $\beta=2$.
This is in excellent agreement with the results from the {\color{black}kinetic theory} of homogeneously driven granular gases \cite{Noije:1997} and also with results from {\color{black}molcular dynamics} simulations \cite{cafiero2002} where $\beta=1.41$ was found.

{\em Conclusion.}
One of the most fundamental characteristics of a homogeneous granular gas is the overpopulation of the high-energy tail of the velocity distribution function. This characteristic distinguishes granular gases from molecular gases which obey a Maxwellian distribution, $P\left(v\right)\sim \exp\left(-v^2\right)$, for large $v$. The overpopulation of the high-energy tail was predicted for both homogeneously cooling and randomly heated gas, leading to high-energy asymptotics, $P\left(v\right)\sim \exp\left(-v^\beta\right)$, with $\beta=1$ in the former case and $\beta=3/2$ in the latter. Both results use the same mathematical apparatus \cite{Esipov:1997,Noije:1997}, thus, they support one another. Despite the fundamental importance of this result evidenced by the large number of papers referring to it, by now, there was no clean experimental confirmation, due to the fact that the available experiments introduce undesired correlations in an uncontrolled way, either by off-plane collisions or surface roughness in vibrated sub-monolayers or by boundary effects in microgravity experiments. {\color{black}In previous experiments, the overpopulation was observed}, however, whether it results from the dissipative collisions in a granular gas or from velocity correlations introduced through the experimental setup, was not free of doubt.

In the experiment presented here, we heat the granular gas by means of the rotational degree of freedom. This idea was introduced by Cafiero et al. \cite{cafiero2002} in the context of numerical simulations. First, we carefully checked that the system fulfills the preconditions under which the theoretical results were obtained, namely (a) a single particle (which cannot dissipatively interact with others) obeys a Gaussian distribution for the linear velocity components, and (b) homogeneity and isotropy of the many-particle system. Once convinced that the preconditions hold true, we measured the velocity distribution function and found an high-energy tail in excellent agreement with the theoretical prediction, $P\left(v\right)\sim \exp\left(-v^{3/2}\right)$.

This experimental result confirms the theoretical prediction and, thus, the {\color{black}kinetic theory} of granular gases being the mathematical apparatus involved in this prediction. This conclusion is not trivial, in our opinion, since the {\color{black}kinetic theory} of granular gases employs several assumptions such as {\color{black}molcular chaos}, the instantaneous collision assumption and others whose validity for granular gases has been questioned.

\medskip
\begin{acknowledgments} 
We acknowledge funding by Deutsche Forschungsgemeinschaft through the Cluster of Excellence ``Engineering of Advanced Materials'', ZISC and FPS and Grant SCHO 1700/1-1.
  We thank Michael Heckel for technical support and Achim Sack and Jonathan Kollmer for discussions. 
\end{acknowledgments} 


\begin{thebibliography}{48}%
\makeatletter
\providecommand \@ifxundefined [1]{%
 \@ifx{#1\undefined}
}%
\providecommand \@ifnum [1]{%
 \ifnum #1\expandafter \@firstoftwo
 \else \expandafter \@secondoftwo
 \fi
}%
\providecommand \@ifx [1]{%
 \ifx #1\expandafter \@firstoftwo
 \else \expandafter \@secondoftwo
 \fi
}%
\providecommand \natexlab [1]{#1}%
\providecommand \enquote  [1]{``#1''}%
\providecommand \bibnamefont  [1]{#1}%
\providecommand \bibfnamefont [1]{#1}%
\providecommand \citenamefont [1]{#1}%
\providecommand \href@noop [0]{\@secondoftwo}%
\providecommand \href [0]{\begingroup \@sanitize@url \@href}%
\providecommand \@href[1]{\@@startlink{#1}\@@href}%
\providecommand \@@href[1]{\endgroup#1\@@endlink}%
\providecommand \@sanitize@url [0]{\catcode `\\12\catcode `\$12\catcode
  `\&12\catcode `\#12\catcode `\^12\catcode `\_12\catcode `\%12\relax}%
\providecommand \@@startlink[1]{}%
\providecommand \@@endlink[0]{}%
\providecommand \url  [0]{\begingroup\@sanitize@url \@url }%
\providecommand \@url [1]{\endgroup\@href {#1}{\urlprefix }}%
\providecommand \urlprefix  [0]{URL }%
\providecommand \Eprint [0]{\href }%
\providecommand \doibase [0]{http://dx.doi.org/}%
\providecommand \selectlanguage [0]{\@gobble}%
\providecommand \bibinfo  [0]{\@secondoftwo}%
\providecommand \bibfield  [0]{\@secondoftwo}%
\providecommand \translation [1]{[#1]}%
\providecommand \BibitemOpen [0]{}%
\providecommand \bibitemStop [0]{}%
\providecommand \bibitemNoStop [0]{.\EOS\space}%
\providecommand \EOS [0]{\spacefactor3000\relax}%
\providecommand \BibitemShut  [1]{\csname bibitem#1\endcsname}%
\let\auto@bib@innerbib\@empty
\bibitem [{\citenamefont {Goldhirsch}(2003)}]{Goldhirsch:2003a}%
  \BibitemOpen
  \bibfield  {author} {\bibinfo {author} {\bibfnamefont {I.}~\bibnamefont
  {Goldhirsch}},\ }\bibfield  {title} {\enquote {\bibinfo {title} {Rapid
  granular flows},}\ }\href@noop {} {\bibfield  {journal} {\bibinfo  {journal}
  {Annu. Rev. Fluid Mech.}\ }\textbf {\bibinfo {volume} {35}},\ \bibinfo
  {pages} {267} (\bibinfo {year} {2003})}\BibitemShut {NoStop}%
\bibitem [{\citenamefont {Spahn}\ and\ \citenamefont
  {Schmidt}(2006)}]{Spahn:2006}%
  \BibitemOpen
  \bibfield  {author} {\bibinfo {author} {\bibfnamefont {F.}~\bibnamefont
  {Spahn}}\ and\ \bibinfo {author} {\bibfnamefont {J.}~\bibnamefont
  {Schmidt}},\ }\bibfield  {title} {\enquote {\bibinfo {title} {Hydrodynamic
  description of planetary rings},}\ }\href@noop {} {\bibfield  {journal}
  {\bibinfo  {journal} {GAMM-Mitt.}\ }\textbf {\bibinfo {volume} {29}},\
  \bibinfo {pages} {115} (\bibinfo {year} {2006})}\BibitemShut {NoStop}%
\bibitem [{\citenamefont {Goldhirsch}\ and\ \citenamefont
  {Zanetti}(1993)}]{Goldhirsch:1993a}%
  \BibitemOpen
  \bibfield  {author} {\bibinfo {author} {\bibfnamefont {I.}~\bibnamefont
  {Goldhirsch}}\ and\ \bibinfo {author} {\bibfnamefont {G.}~\bibnamefont
  {Zanetti}},\ }\bibfield  {title} {\enquote {\bibinfo {title} {Clustering
  instability in dissipative gases},}\ }\href@noop {} {\bibfield  {journal}
  {\bibinfo  {journal} {Phys. Rev. Lett.}\ }\textbf {\bibinfo {volume} {70}},\
  \bibinfo {pages} {1619--22} (\bibinfo {year} {1993})}\BibitemShut {NoStop}%
\bibitem [{\citenamefont {van Noije}\ \emph {et~al.}(1997)\citenamefont {van
  Noije}, \citenamefont {Ernst}, \citenamefont {Brito},\ and\ \citenamefont
  {Orza}}]{Noije:1997}%
  \BibitemOpen
  \bibfield  {author} {\bibinfo {author} {\bibfnamefont {T.~P.~C.}\
  \bibnamefont {van Noije}}, \bibinfo {author} {\bibfnamefont {M.~H.}\
  \bibnamefont {Ernst}}, \bibinfo {author} {\bibfnamefont {R.}~\bibnamefont
  {Brito}}, \ and\ \bibinfo {author} {\bibfnamefont {J.~A.~G.}\ \bibnamefont
  {Orza}},\ }\bibfield  {title} {\enquote {\bibinfo {title} {Mesoscopic theory
  of granular fluids},}\ }\href@noop {} {\bibfield  {journal} {\bibinfo
  {journal} {Phys. Rev. Lett.}\ }\textbf {\bibinfo {volume} {79}},\ \bibinfo
  {pages} {411} (\bibinfo {year} {1997})}\BibitemShut {NoStop}%
\bibitem [{\citenamefont {Brey}\ \emph {et~al.}(1997)\citenamefont {Brey},
  \citenamefont {Dufty},\ and\ \citenamefont {Santos}}]{Brey:1997}%
  \BibitemOpen
  \bibfield  {author} {\bibinfo {author} {\bibfnamefont {J.~J.}\ \bibnamefont
  {Brey}}, \bibinfo {author} {\bibfnamefont {J.~W.}\ \bibnamefont {Dufty}}, \
  and\ \bibinfo {author} {\bibfnamefont {A.}~\bibnamefont {Santos}},\
  }\bibfield  {title} {\enquote {\bibinfo {title} {Dissipative dynamics for
  hard spheres},}\ }\href@noop {} {\bibfield  {journal} {\bibinfo  {journal}
  {J. Stat. Phys.}\ }\textbf {\bibinfo {volume} {87}},\ \bibinfo {pages} {1051}
  (\bibinfo {year} {1997})}\BibitemShut {NoStop}%
\bibitem [{\citenamefont {van Noije}\ \emph {et~al.}(1998)\citenamefont {van
  Noije}, \citenamefont {Ernst},\ and\ \citenamefont {Brito}}]{Noije:1997b}%
  \BibitemOpen
  \bibfield  {author} {\bibinfo {author} {\bibfnamefont {T.~P.~C.}\
  \bibnamefont {van Noije}}, \bibinfo {author} {\bibfnamefont {M.~H.}\
  \bibnamefont {Ernst}}, \ and\ \bibinfo {author} {\bibfnamefont
  {R.}~\bibnamefont {Brito}},\ }\bibfield  {title} {\enquote {\bibinfo {title}
  {Ring kinetic theory for an idealized granular gas},}\ }\href@noop {}
  {\bibfield  {journal} {\bibinfo  {journal} {Physica A}\ }\textbf {\bibinfo
  {volume} {251}},\ \bibinfo {pages} {266} (\bibinfo {year}
  {1998})}\BibitemShut {NoStop}%
\bibitem [{\citenamefont {Brilliantov}\ and\ \citenamefont
  {P\"oschel}(2004)}]{SMGG}%
  \BibitemOpen
  \bibfield  {author} {\bibinfo {author} {\bibfnamefont {N.~V.}\ \bibnamefont
  {Brilliantov}}\ and\ \bibinfo {author} {\bibfnamefont {T.}~\bibnamefont
  {P\"oschel}},\ }\href@noop {} {\emph {\bibinfo {title} {Kinetic Theory of
  Granular Gases}}}\ (\bibinfo  {publisher} {Oxford University Press},\
  \bibinfo {year} {2004})\BibitemShut {NoStop}%
\bibitem [{\citenamefont {Sela}\ and\ \citenamefont
  {Goldhirsch}(1998)}]{Sela:1998}%
  \BibitemOpen
  \bibfield  {author} {\bibinfo {author} {\bibfnamefont {N.}~\bibnamefont
  {Sela}}\ and\ \bibinfo {author} {\bibfnamefont {I.}~\bibnamefont
  {Goldhirsch}},\ }\bibfield  {title} {\enquote {\bibinfo {title} {Hydrodynamic
  equations for rapid flows of smooth inelastic spheres, to {B}urnett order},}\
  }\href@noop {} {\bibfield  {journal} {\bibinfo  {journal} {J. Fluid Mech.}\
  }\textbf {\bibinfo {volume} {361}},\ \bibinfo {pages} {41} (\bibinfo {year}
  {1998})}\BibitemShut {NoStop}%
\bibitem [{\citenamefont {Brey}\ \emph {et~al.}(1998)\citenamefont {Brey},
  \citenamefont {Dufty}, \citenamefont {Kim},\ and\ \citenamefont
  {Santos}}]{Brey:1998}%
  \BibitemOpen
  \bibfield  {author} {\bibinfo {author} {\bibfnamefont {J.~J.}\ \bibnamefont
  {Brey}}, \bibinfo {author} {\bibfnamefont {J.~W.}\ \bibnamefont {Dufty}},
  \bibinfo {author} {\bibfnamefont {C.~S.}\ \bibnamefont {Kim}}, \ and\
  \bibinfo {author} {\bibfnamefont {A.}~\bibnamefont {Santos}},\ }\bibfield
  {title} {\enquote {\bibinfo {title} {Hydrodynamics for granular flow at low
  density},}\ }\href@noop {} {\bibfield  {journal} {\bibinfo  {journal} {Phys.
  Rev. E}\ }\textbf {\bibinfo {volume} {58}},\ \bibinfo {pages} {4638}
  (\bibinfo {year} {1998})}\BibitemShut {NoStop}%
\bibitem [{\citenamefont {Brilliantov}\ and\ \citenamefont
  {P\"oschel}(2003)}]{Brilliantov:2003}%
  \BibitemOpen
  \bibfield  {author} {\bibinfo {author} {\bibfnamefont {N.~V.}\ \bibnamefont
  {Brilliantov}}\ and\ \bibinfo {author} {\bibfnamefont {T.}~\bibnamefont
  {P\"oschel}},\ }\bibfield  {title} {\enquote {\bibinfo {title} {Hydrodynamics
  and transport coefficients for granular gases},}\ }\href@noop {} {\bibfield
  {journal} {\bibinfo  {journal} {Phys. Rev. E}\ }\textbf {\bibinfo {volume}
  {67}},\ \bibinfo {pages} {061304} (\bibinfo {year} {2003})}\BibitemShut
  {NoStop}%
\bibitem [{\citenamefont {Brilliantov}\ and\ \citenamefont
  {P\"oschel}(2000)}]{Brilliantov:2000a}%
  \BibitemOpen
  \bibfield  {author} {\bibinfo {author} {\bibfnamefont {N.~V.}\ \bibnamefont
  {Brilliantov}}\ and\ \bibinfo {author} {\bibfnamefont {T.}~\bibnamefont
  {P\"oschel}},\ }\bibfield  {title} {\enquote {\bibinfo {title}
  {Self-diffusion in granular gases},}\ }\href@noop {} {\bibfield  {journal}
  {\bibinfo  {journal} {Phys. Rev. E}\ }\textbf {\bibinfo {volume} {61}},\
  \bibinfo {pages} {1716} (\bibinfo {year} {2000})}\BibitemShut {NoStop}%
\bibitem [{\citenamefont {Brilliantov}\ and\ \citenamefont
  {P\"oschel}(2005)}]{Brilliantov:2005}%
  \BibitemOpen
  \bibfield  {author} {\bibinfo {author} {\bibfnamefont {N.~V.}\ \bibnamefont
  {Brilliantov}}\ and\ \bibinfo {author} {\bibfnamefont {T.}~\bibnamefont
  {P\"oschel}},\ }\bibfield  {title} {\enquote {\bibinfo {title}
  {Self-diffusion in granular gases: {G}reen-{K}ubo versus
  {C}hapman-{E}nskog},}\ }\href@noop {} {\bibfield  {journal} {\bibinfo
  {journal} {Chaos}\ }\textbf {\bibinfo {volume} {15}},\ \bibinfo {pages}
  {026108} (\bibinfo {year} {2005})}\BibitemShut {NoStop}%
\bibitem [{\citenamefont {Brilliantov}\ \emph {et~al.}(2007)\citenamefont
  {Brilliantov}, \citenamefont {P\"oschel}, \citenamefont {Kranz},\ and\
  \citenamefont {Zippelius}}]{Brilliantov:2007a}%
  \BibitemOpen
  \bibfield  {author} {\bibinfo {author} {\bibfnamefont {N.~V.}\ \bibnamefont
  {Brilliantov}}, \bibinfo {author} {\bibfnamefont {T.}~\bibnamefont
  {P\"oschel}}, \bibinfo {author} {\bibfnamefont {W.~T.}\ \bibnamefont
  {Kranz}}, \ and\ \bibinfo {author} {\bibfnamefont {A.}~\bibnamefont
  {Zippelius}},\ }\bibfield  {title} {\enquote {\bibinfo {title} {Translations
  and rotations are correlated in granular gases},}\ }\href@noop {} {\bibfield
  {journal} {\bibinfo  {journal} {Phys. Rev. Lett.}\ }\textbf {\bibinfo
  {volume} {98}},\ \bibinfo {pages} {128001} (\bibinfo {year}
  {2007})}\BibitemShut {NoStop}%
\bibitem [{\citenamefont {Kranz}\ \emph {et~al.}(2009)\citenamefont {Kranz},
  \citenamefont {Brilliantov}, \citenamefont {P{\"o}schel},\ and\ \citenamefont
  {Zippelius}}]{Kranz:2009}%
  \BibitemOpen
  \bibfield  {author} {\bibinfo {author} {\bibfnamefont {W.~T.}\ \bibnamefont
  {Kranz}}, \bibinfo {author} {\bibfnamefont {N.~V.}\ \bibnamefont
  {Brilliantov}}, \bibinfo {author} {\bibfnamefont {T.}~\bibnamefont
  {P{\"o}schel}}, \ and\ \bibinfo {author} {\bibfnamefont {A.}~\bibnamefont
  {Zippelius}},\ }\bibfield  {title} {\enquote {\bibinfo {title} {Correlation
  of spin and velocity in the homogeneous cooling state of a granular gas of
  rough particles},}\ }\href@noop {} {\bibfield  {journal} {\bibinfo  {journal}
  {The European Physical Journal Special Topics}\ }\textbf {\bibinfo {volume}
  {179}},\ \bibinfo {pages} {91} (\bibinfo {year} {2009})}\BibitemShut
  {NoStop}%
\bibitem [{\citenamefont {Goldshtein}\ and\ \citenamefont
  {Shapiro}(1995)}]{Goldshtein:1995}%
  \BibitemOpen
  \bibfield  {author} {\bibinfo {author} {\bibfnamefont {A.}~\bibnamefont
  {Goldshtein}}\ and\ \bibinfo {author} {\bibfnamefont {M.}~\bibnamefont
  {Shapiro}},\ }\bibfield  {title} {\enquote {\bibinfo {title} {Mechanics of
  collisional motion of granular materials. part 1. general hydrodynamic
  equations},}\ }\href@noop {} {\bibfield  {journal} {\bibinfo  {journal} {J.
  Fluid Mech.}\ }\textbf {\bibinfo {volume} {282}},\ \bibinfo {pages} {75}
  (\bibinfo {year} {1995})}\BibitemShut {NoStop}%
\bibitem [{\citenamefont {Esipov}\ and\ \citenamefont
  {P\"oschel}(1997)}]{Esipov:1997}%
  \BibitemOpen
  \bibfield  {author} {\bibinfo {author} {\bibfnamefont {S.~E.}\ \bibnamefont
  {Esipov}}\ and\ \bibinfo {author} {\bibfnamefont {T.}~\bibnamefont
  {P\"oschel}},\ }\bibfield  {title} {\enquote {\bibinfo {title} {The granular
  phase diagram},}\ }\href@noop {} {\bibfield  {journal} {\bibinfo  {journal}
  {J. Stat. Phys.}\ }\textbf {\bibinfo {volume} {86}},\ \bibinfo {pages}
  {1385--1395} (\bibinfo {year} {1997})}\BibitemShut {NoStop}%
\bibitem [{\citenamefont {P{\"o}schel}\ \emph {et~al.}(2006)\citenamefont
  {P{\"o}schel}, \citenamefont {Brilliantov},\ and\ \citenamefont
  {Formella}}]{Poeschel:2006}%
  \BibitemOpen
  \bibfield  {author} {\bibinfo {author} {\bibfnamefont {T.}~\bibnamefont
  {P{\"o}schel}}, \bibinfo {author} {\bibfnamefont {N.~V.}\ \bibnamefont
  {Brilliantov}}, \ and\ \bibinfo {author} {\bibfnamefont {A.}~\bibnamefont
  {Formella}},\ }\bibfield  {title} {\enquote {\bibinfo {title} {Impact of
  high-energy tails on granular gas properties},}\ }\href@noop {} {\bibfield
  {journal} {\bibinfo  {journal} {Phys. Rev. E}\ }\textbf {\bibinfo {volume}
  {74}},\ \bibinfo {pages} {041302} (\bibinfo {year} {2006})}\BibitemShut
  {NoStop}%
\bibitem [{\citenamefont {P\"oschel}\ \emph {et~al.}(2007)\citenamefont
  {P\"oschel}, \citenamefont {Brilliantov},\ and\ \citenamefont
  {Formella}}]{Poeschel:2007}%
  \BibitemOpen
  \bibfield  {author} {\bibinfo {author} {\bibfnamefont {T.}~\bibnamefont
  {P\"oschel}}, \bibinfo {author} {\bibfnamefont {N.~V.}\ \bibnamefont
  {Brilliantov}}, \ and\ \bibinfo {author} {\bibfnamefont {A.}~\bibnamefont
  {Formella}},\ }\bibfield  {title} {\enquote {\bibinfo {title} {Granular gas
  cooling and relaxation to the steady state in regard to the overpopulated
  tail of the velocity distribution},}\ }\href@noop {} {\bibfield  {journal}
  {\bibinfo  {journal} {Int. J. Mod. Phys. C}\ }\textbf {\bibinfo {volume}
  {18}},\ \bibinfo {pages} {701} (\bibinfo {year} {2007})}\BibitemShut
  {NoStop}%
\bibitem [{\citenamefont {Goldhirsch}\ \emph {et~al.}(2003)\citenamefont
  {Goldhirsch}, \citenamefont {Noskowicz},\ and\ \citenamefont
  {Bar-Lev}}]{Goldhirsch:2003b}%
  \BibitemOpen
  \bibfield  {author} {\bibinfo {author} {\bibfnamefont {I.}~\bibnamefont
  {Goldhirsch}}, \bibinfo {author} {\bibfnamefont {S.~H.}\ \bibnamefont
  {Noskowicz}}, \ and\ \bibinfo {author} {\bibfnamefont {O.}~\bibnamefont
  {Bar-Lev}},\ }\bibfield  {title} {\enquote {\bibinfo {title} {The homogeneous
  cooling state revisited},}\ }\href@noop {} {\bibfield  {journal} {\bibinfo
  {journal} {Lecture Notes in Physics}\ }\textbf {\bibinfo {volume} {624}},\
  \bibinfo {pages} {37--63} (\bibinfo {year} {2003})}\BibitemShut {NoStop}%
\bibitem [{\citenamefont {Noskowicz}\ \emph {et~al.}(2007)\citenamefont
  {Noskowicz}, \citenamefont {Bar-Lev},\ and\ \citenamefont
  {Serero}}]{Noskowicz:2007}%
  \BibitemOpen
  \bibfield  {author} {\bibinfo {author} {\bibfnamefont {S.~H.}\ \bibnamefont
  {Noskowicz}}, \bibinfo {author} {\bibfnamefont {O.}~\bibnamefont {Bar-Lev}},
  \ and\ \bibinfo {author} {\bibfnamefont {D.}~\bibnamefont {Serero}},\
  }\bibfield  {title} {\enquote {\bibinfo {title} {Sonine},}\ }\href@noop {}
  {\bibfield  {journal} {\bibinfo  {journal} {Europhys. Lett.}\ }\textbf
  {\bibinfo {volume} {79}},\ \bibinfo {pages} {60001} (\bibinfo {year}
  {2007})}\BibitemShut {NoStop}%
\bibitem [{\citenamefont {Villani}(2006)}]{Villani:2006}%
  \BibitemOpen
  \bibfield  {author} {\bibinfo {author} {\bibfnamefont {C.}~\bibnamefont
  {Villani}},\ }\bibfield  {title} {\enquote {\bibinfo {title} {Mathematics of
  granular materials},}\ }\href@noop {} {\bibfield  {journal} {\bibinfo
  {journal} {J. Stat. Phys.}\ }\textbf {\bibinfo {volume} {124}},\ \bibinfo
  {pages} {781} (\bibinfo {year} {2006})}\BibitemShut {NoStop}%
\bibitem [{\citenamefont {Brey}\ and\ \citenamefont
  {Ruiz-Montero}(1999)}]{Brey:1999}%
  \BibitemOpen
  \bibfield  {author} {\bibinfo {author} {\bibfnamefont {J.~J.}\ \bibnamefont
  {Brey}}\ and\ \bibinfo {author} {\bibfnamefont {M.~J.}\ \bibnamefont
  {Ruiz-Montero}},\ }\bibfield  {title} {\enquote {\bibinfo {title} {Direct
  monte carlo simulation of dilute granular flow},}\ }\href@noop {} {\bibfield
  {journal} {\bibinfo  {journal} {Comp. Phys. Comm.}\ }\textbf {\bibinfo
  {volume} {121}},\ \bibinfo {pages} {278} (\bibinfo {year}
  {1999})}\BibitemShut {NoStop}%
\bibitem [{\citenamefont {Puglisi}\ \emph {et~al.}(1999)\citenamefont
  {Puglisi}, \citenamefont {Loreto}, \citenamefont {Marini Bettolo~Marconi},\
  and\ \citenamefont {Vulpiani}}]{Puglisi:1999}%
  \BibitemOpen
  \bibfield  {author} {\bibinfo {author} {\bibfnamefont {A.}~\bibnamefont
  {Puglisi}}, \bibinfo {author} {\bibfnamefont {V.}~\bibnamefont {Loreto}},
  \bibinfo {author} {\bibfnamefont {U.}~\bibnamefont {Marini Bettolo~Marconi}},
  \ and\ \bibinfo {author} {\bibfnamefont {A.}~\bibnamefont {Vulpiani}},\
  }\bibfield  {title} {\enquote {\bibinfo {title} {Kinetic approach to granular
  gases},}\ }\href@noop {} {\bibfield  {journal} {\bibinfo  {journal} {Phys.
  Rev. E}\ }\textbf {\bibinfo {volume} {59}},\ \bibinfo {pages} {5582}
  (\bibinfo {year} {1999})}\BibitemShut {NoStop}%
\bibitem [{\citenamefont {Moon}\ \emph {et~al.}(2004)\citenamefont {Moon},
  \citenamefont {Swift},\ and\ \citenamefont {Swinney}}]{Moon:2003}%
  \BibitemOpen
  \bibfield  {author} {\bibinfo {author} {\bibfnamefont {S.~J.}\ \bibnamefont
  {Moon}}, \bibinfo {author} {\bibfnamefont {J.~B.}\ \bibnamefont {Swift}}, \
  and\ \bibinfo {author} {\bibfnamefont {H.~L.}\ \bibnamefont {Swinney}},\
  }\bibfield  {title} {\enquote {\bibinfo {title} {Steady-state velocity
  distributions of an oscillated granular gas steady-state velocity
  distributions of an oscillated granular gas steady-state velocity
  distribution of an oscillated granular gas},}\ }\href@noop {} {\bibfield
  {journal} {\bibinfo  {journal} {Phys. Rev. E}\ }\textbf {\bibinfo {volume}
  {69}},\ \bibinfo {pages} {011301} (\bibinfo {year} {2004})}\BibitemShut
  {NoStop}%
\bibitem [{\citenamefont {Brey}\ \emph {et~al.}(1999)\citenamefont {Brey},
  \citenamefont {Cubero},\ and\ \citenamefont {Ruiz-Montero}}]{Brey:1999a}%
  \BibitemOpen
  \bibfield  {author} {\bibinfo {author} {\bibfnamefont {J.~J.}\ \bibnamefont
  {Brey}}, \bibinfo {author} {\bibfnamefont {D.}~\bibnamefont {Cubero}}, \ and\
  \bibinfo {author} {\bibfnamefont {M.~J.}\ \bibnamefont {Ruiz-Montero}},\
  }\bibfield  {title} {\enquote {\bibinfo {title} {High energy tail in the
  velocity distribution of a granular gas},}\ }\href@noop {} {\bibfield
  {journal} {\bibinfo  {journal} {Phys. Rev. E}\ }\textbf {\bibinfo {volume}
  {59}},\ \bibinfo {pages} {1256} (\bibinfo {year} {1999})}\BibitemShut
  {NoStop}%
\bibitem [{\citenamefont {van Zon}\ and\ \citenamefont
  {MacKintosh}(2005)}]{vanZorn:2005}%
  \BibitemOpen
  \bibfield  {author} {\bibinfo {author} {\bibfnamefont {J.~S.}\ \bibnamefont
  {van Zon}}\ and\ \bibinfo {author} {\bibfnamefont {F.~C.}\ \bibnamefont
  {MacKintosh}},\ }\bibfield  {title} {\enquote {\bibinfo {title} {Velocity
  distributions in dilute granular systems},}\ }\href@noop {} {\bibfield
  {journal} {\bibinfo  {journal} {Phys. Rev. E}\ }\textbf {\bibinfo {volume}
  {72}},\ \bibinfo {pages} {051301} (\bibinfo {year} {2005})}\BibitemShut
  {NoStop}%
\bibitem [{\citenamefont {Meerson}\ \emph {et~al.}(2004)\citenamefont
  {Meerson}, \citenamefont {P\"oschel}, \citenamefont {Sasorov},\ and\
  \citenamefont {Schwager}}]{Meerson:2004}%
  \BibitemOpen
  \bibfield  {author} {\bibinfo {author} {\bibfnamefont {B.}~\bibnamefont
  {Meerson}}, \bibinfo {author} {\bibfnamefont {T.}~\bibnamefont {P\"oschel}},
  \bibinfo {author} {\bibfnamefont {P.~V.}\ \bibnamefont {Sasorov}}, \ and\
  \bibinfo {author} {\bibfnamefont {T.}~\bibnamefont {Schwager}},\ }\bibfield
  {title} {\enquote {\bibinfo {title} {Giant fluctuations at a granular phase
  separation threshold},}\ }\href@noop {} {\bibfield  {journal} {\bibinfo
  {journal} {Phys. Rev. E}\ }\textbf {\bibinfo {volume} {69}},\ \bibinfo
  {pages} {021302} (\bibinfo {year} {2004})}\BibitemShut {NoStop}%
\bibitem [{\citenamefont {Grasselli}\ \emph {et~al.}(2015)\citenamefont
  {Grasselli}, \citenamefont {Bossis},\ and\ \citenamefont
  {Morini}}]{Grasselli:2015}%
  \BibitemOpen
  \bibfield  {author} {\bibinfo {author} {\bibfnamefont {Y.}~\bibnamefont
  {Grasselli}}, \bibinfo {author} {\bibfnamefont {G.}~\bibnamefont {Bossis}}, \
  and\ \bibinfo {author} {\bibfnamefont {R.}~\bibnamefont {Morini}},\
  }\bibfield  {title} {\enquote {\bibinfo {title} {Translational and rotational
  temperatures of a 2d vibrated granular gas in microgravity},}\ }\href@noop {}
  {\bibfield  {journal} {\bibinfo  {journal} {Eur. Phys. J. E}\ }\textbf
  {\bibinfo {volume} {38}},\ \bibinfo {pages} {1} (\bibinfo {year}
  {2015})}\BibitemShut {NoStop}%
\bibitem [{\citenamefont {Tatsumi}\ \emph {et~al.}(2009)\citenamefont
  {Tatsumi}, \citenamefont {Murayama}, \citenamefont {Hayakawa},\ and\
  \citenamefont {Masaki}}]{Tatsumi:2009}%
  \BibitemOpen
  \bibfield  {author} {\bibinfo {author} {\bibfnamefont {S.}~\bibnamefont
  {Tatsumi}}, \bibinfo {author} {\bibfnamefont {Y.}~\bibnamefont {Murayama}},
  \bibinfo {author} {\bibfnamefont {H.}~\bibnamefont {Hayakawa}}, \ and\
  \bibinfo {author} {\bibfnamefont {S.}~\bibnamefont {Masaki}},\ }\bibfield
  {title} {\enquote {\bibinfo {title} {Experimental study on the kinetics of
  granular gases under microgravity},}\ }\href@noop {} {\bibfield  {journal}
  {\bibinfo  {journal} {Journal of Fluid Mechanics}\ }\textbf {\bibinfo
  {volume} {641}},\ \bibinfo {pages} {521} (\bibinfo {year}
  {2009})}\BibitemShut {NoStop}%
\bibitem [{\citenamefont {Hou}\ \emph {et~al.}(2008)\citenamefont {Hou},
  \citenamefont {Liu}, \citenamefont {Zhai}, \citenamefont {Sun}, \citenamefont
  {Lu}, \citenamefont {Garrabos},\ and\ \citenamefont {Evesque}}]{Hou:2008}%
  \BibitemOpen
  \bibfield  {author} {\bibinfo {author} {\bibfnamefont {M.}~\bibnamefont
  {Hou}}, \bibinfo {author} {\bibfnamefont {R.}~\bibnamefont {Liu}}, \bibinfo
  {author} {\bibfnamefont {G.}~\bibnamefont {Zhai}}, \bibinfo {author}
  {\bibfnamefont {Z.}~\bibnamefont {Sun}}, \bibinfo {author} {\bibfnamefont
  {K.}~\bibnamefont {Lu}}, \bibinfo {author} {\bibfnamefont {Y.}~\bibnamefont
  {Garrabos}}, \ and\ \bibinfo {author} {\bibfnamefont {P.}~\bibnamefont
  {Evesque}},\ }\bibfield  {title} {\enquote {\bibinfo {title} {Velocity
  distribution of vibration-driven granular gas in knudsen regime in
  microgravity},}\ }\href@noop {} {\bibfield  {journal} {\bibinfo  {journal}
  {Microgravity Science and Technology}\ }\textbf {\bibinfo {volume} {20}},\
  \bibinfo {pages} {73} (\bibinfo {year} {2008})}\BibitemShut {NoStop}%
\bibitem [{\citenamefont {Van~Noije}\ and\ \citenamefont
  {Ernst}(1998)}]{van1998velocity}%
  \BibitemOpen
  \bibfield  {author} {\bibinfo {author} {\bibfnamefont {T.~P.~C.}\
  \bibnamefont {Van~Noije}}\ and\ \bibinfo {author} {\bibfnamefont {M.~H.}\
  \bibnamefont {Ernst}},\ }\bibfield  {title} {\enquote {\bibinfo {title}
  {Velocity distributions in homogeneous granular fluids: the free and the
  heated case},}\ }\href@noop {} {\bibfield  {journal} {\bibinfo  {journal}
  {Granul.~Matter}\ }\textbf {\bibinfo {volume} {1}},\ \bibinfo {pages}
  {57--64} (\bibinfo {year} {1998})}\BibitemShut {NoStop}%
\bibitem [{\citenamefont {Schmick}\ and\ \citenamefont
  {Markus}(2008)}]{Schmick:2008}%
  \BibitemOpen
  \bibfield  {author} {\bibinfo {author} {\bibfnamefont {M.}~\bibnamefont
  {Schmick}}\ and\ \bibinfo {author} {\bibfnamefont {M.}~\bibnamefont
  {Markus}},\ }\bibfield  {title} {\enquote {\bibinfo {title} {Gaussian
  distributions of rotational velocities in a granular medium},}\ }\href@noop
  {} {\bibfield  {journal} {\bibinfo  {journal} {Phys. Rev. E}\ }\textbf
  {\bibinfo {volume} {78}},\ \bibinfo {pages} {010302} (\bibinfo {year}
  {2008})}\BibitemShut {NoStop}%
\bibitem [{\citenamefont {Rouyer}\ and\ \citenamefont
  {Menon}(2000)}]{Rouyer:2000}%
  \BibitemOpen
  \bibfield  {author} {\bibinfo {author} {\bibfnamefont {F.}~\bibnamefont
  {Rouyer}}\ and\ \bibinfo {author} {\bibfnamefont {N.}~\bibnamefont {Menon}},\
  }\bibfield  {title} {\enquote {\bibinfo {title} {Velocity fluctuations in a
  homogeneous 2d granular gas in steady state},}\ }\href@noop {} {\bibfield
  {journal} {\bibinfo  {journal} {Phys. Rev. Lett.}\ }\textbf {\bibinfo
  {volume} {85}},\ \bibinfo {pages} {3676} (\bibinfo {year}
  {2000})}\BibitemShut {NoStop}%
\bibitem [{\citenamefont {Baxter}\ and\ \citenamefont
  {Olafsen}(2003)}]{Baxter:2003}%
  \BibitemOpen
  \bibfield  {author} {\bibinfo {author} {\bibfnamefont {G.~W.}\ \bibnamefont
  {Baxter}}\ and\ \bibinfo {author} {\bibfnamefont {J.~S.}\ \bibnamefont
  {Olafsen}},\ }\bibfield  {title} {\enquote {\bibinfo {title} {Kinetics:
  Gaussian statistics in granular gases},}\ }\href@noop {} {\bibfield
  {journal} {\bibinfo  {journal} {Nature}\ }\textbf {\bibinfo {volume} {425}},\
  \bibinfo {pages} {680} (\bibinfo {year} {2003})}\BibitemShut {NoStop}%
\bibitem [{\citenamefont {Olafsen}\ and\ \citenamefont
  {Urbach}(1999)}]{Olafsen:1999}%
  \BibitemOpen
  \bibfield  {author} {\bibinfo {author} {\bibfnamefont {J.~S.}\ \bibnamefont
  {Olafsen}}\ and\ \bibinfo {author} {\bibfnamefont {J.~S.}\ \bibnamefont
  {Urbach}},\ }\bibfield  {title} {\enquote {\bibinfo {title} {Velocity
  distributions and density fluctuations in a granular gas},}\ }\href@noop {}
  {\bibfield  {journal} {\bibinfo  {journal} {Phys. Rev. E}\ }\textbf {\bibinfo
  {volume} {60}},\ \bibinfo {pages} {R2468} (\bibinfo {year}
  {1999})}\BibitemShut {NoStop}%
\bibitem [{\citenamefont {Losert}\ \emph {et~al.}(1999)\citenamefont {Losert},
  \citenamefont {Cooper}, \citenamefont {Delour}, \citenamefont {Kudrolli},\
  and\ \citenamefont {Gollub}}]{Losert:1999}%
  \BibitemOpen
  \bibfield  {author} {\bibinfo {author} {\bibfnamefont {W.}~\bibnamefont
  {Losert}}, \bibinfo {author} {\bibfnamefont {D.~G.~W.}\ \bibnamefont
  {Cooper}}, \bibinfo {author} {\bibfnamefont {J.}~\bibnamefont {Delour}},
  \bibinfo {author} {\bibfnamefont {A.}~\bibnamefont {Kudrolli}}, \ and\
  \bibinfo {author} {\bibfnamefont {J.~P.}\ \bibnamefont {Gollub}},\ }\bibfield
   {title} {\enquote {\bibinfo {title} {Velocity statistics in excited granular
  media},}\ }\href@noop {} {\bibfield  {journal} {\bibinfo  {journal} {Chaos}\
  }\textbf {\bibinfo {volume} {9}},\ \bibinfo {pages} {682} (\bibinfo {year}
  {1999})}\BibitemShut {NoStop}%
\bibitem [{\citenamefont {Kudrolli}\ and\ \citenamefont
  {Henry}(2000)}]{Kudrolli:2000}%
  \BibitemOpen
  \bibfield  {author} {\bibinfo {author} {\bibfnamefont {A.}~\bibnamefont
  {Kudrolli}}\ and\ \bibinfo {author} {\bibfnamefont {J.}~\bibnamefont
  {Henry}},\ }\bibfield  {title} {\enquote {\bibinfo {title} {Non-gaussian
  velocity distributions in excited granular matter in the absence of
  clustering},}\ }\href@noop {} {\bibfield  {journal} {\bibinfo  {journal}
  {Phys. Rev. E}\ }\textbf {\bibinfo {volume} {62}},\ \bibinfo {pages} {R1489}
  (\bibinfo {year} {2000})}\BibitemShut {NoStop}%
\bibitem [{\citenamefont {Cafiero}\ \emph {et~al.}(2002)\citenamefont
  {Cafiero}, \citenamefont {Luding},\ and\ \citenamefont
  {Herrmann}}]{cafiero2002}%
  \BibitemOpen
  \bibfield  {author} {\bibinfo {author} {\bibfnamefont {R.}~\bibnamefont
  {Cafiero}}, \bibinfo {author} {\bibfnamefont {S.}~\bibnamefont {Luding}}, \
  and\ \bibinfo {author} {\bibfnamefont {H.~J.}\ \bibnamefont {Herrmann}},\
  }\bibfield  {title} {\enquote {\bibinfo {title} {Rotationally driven gas of
  inelastic rough spheres},}\ }\href@noop {} {\bibfield  {journal} {\bibinfo
  {journal} {Europhys. Lett.}\ }\textbf {\bibinfo {volume} {60}},\ \bibinfo
  {pages} {854} (\bibinfo {year} {2002})}\BibitemShut {NoStop}%
\bibitem [{\citenamefont {Scholz}\ \emph {et~al.}(2016)\citenamefont {Scholz},
  \citenamefont {D’Silva},\ and\ \citenamefont {P{\"o}schel}}]{Scholz:2016}%
  \BibitemOpen
  \bibfield  {author} {\bibinfo {author} {\bibfnamefont {Christian}\
  \bibnamefont {Scholz}}, \bibinfo {author} {\bibfnamefont {Sean}\ \bibnamefont
  {D’Silva}}, \ and\ \bibinfo {author} {\bibfnamefont {Thorsten}\
  \bibnamefont {P{\"o}schel}},\ }\bibfield  {title} {\enquote {\bibinfo {title}
  {Ratcheting and tumbling motion of vibrots},}\ }\href@noop {} {\bibfield
  {journal} {\bibinfo  {journal} {New J. Phys.}\ }\textbf {\bibinfo {volume}
  {18}},\ \bibinfo {pages} {123001} (\bibinfo {year} {2016})}\BibitemShut
  {NoStop}%
\bibitem [{\citenamefont {Altshuler}\ \emph {et~al.}(2013)\citenamefont
  {Altshuler}, \citenamefont {Pastor}, \citenamefont {Garcimart{\'\i}n},
  \citenamefont {Zuriguel},\ and\ \citenamefont {Maza}}]{altshuler2013}%
  \BibitemOpen
  \bibfield  {author} {\bibinfo {author} {\bibfnamefont {E.}~\bibnamefont
  {Altshuler}}, \bibinfo {author} {\bibfnamefont {J.~M.}\ \bibnamefont
  {Pastor}}, \bibinfo {author} {\bibfnamefont {A.}~\bibnamefont
  {Garcimart{\'\i}n}}, \bibinfo {author} {\bibfnamefont {I.}~\bibnamefont
  {Zuriguel}}, \ and\ \bibinfo {author} {\bibfnamefont {D.}~\bibnamefont
  {Maza}},\ }\bibfield  {title} {\enquote {\bibinfo {title} {Vibrot, a simple
  device for the conversion of vibration into rotation mediated by friction:
  preliminary evaluation},}\ }\href@noop {} {\bibfield  {journal} {\bibinfo
  {journal} {PloS one}\ }\textbf {\bibinfo {volume} {8}},\ \bibinfo {pages}
  {e67838} (\bibinfo {year} {2013})}\BibitemShut {NoStop}%
\bibitem [{\citenamefont {Scholz}\ and\ \citenamefont
  {P{\"o}schel}(2016)}]{Scholz2015}%
  \BibitemOpen
  \bibfield  {author} {\bibinfo {author} {\bibfnamefont {C.}~\bibnamefont
  {Scholz}}\ and\ \bibinfo {author} {\bibfnamefont {T.}~\bibnamefont
  {P{\"o}schel}},\ }\bibfield  {title} {\enquote {\bibinfo {title} {Actively
  rotating granular particles manufactured by rapid prototyping},}\ }\href@noop
  {} {\bibfield  {journal} {\bibinfo  {journal} {Rev. Cuba. F{\'\i}sica}\ }
  (\bibinfo {year} {2016})}\BibitemShut {NoStop}%
\bibitem [{\citenamefont {Deseigne}\ \emph {et~al.}(2010)\citenamefont
  {Deseigne}, \citenamefont {Dauchot},\ and\ \citenamefont
  {Chat{\'e}}}]{deseigne2010}%
  \BibitemOpen
  \bibfield  {author} {\bibinfo {author} {\bibfnamefont {J.}~\bibnamefont
  {Deseigne}}, \bibinfo {author} {\bibfnamefont {O.}~\bibnamefont {Dauchot}}, \
  and\ \bibinfo {author} {\bibfnamefont {H.}~\bibnamefont {Chat{\'e}}},\
  }\bibfield  {title} {\enquote {\bibinfo {title} {Collective motion of
  vibrated polar disks},}\ }\href@noop {} {\bibfield  {journal} {\bibinfo
  {journal} {Phys.~Rev.~Lett.}\ }\textbf {\bibinfo {volume} {105}},\ \bibinfo
  {pages} {098001} (\bibinfo {year} {2010})}\BibitemShut {NoStop}%
\bibitem [{\citenamefont {Deseigne}\ \emph {et~al.}(2012)\citenamefont
  {Deseigne}, \citenamefont {L{\'e}onard}, \citenamefont {Dauchot},\ and\
  \citenamefont {Chat{\'e}}}]{deseigne2012vibrated}%
  \BibitemOpen
  \bibfield  {author} {\bibinfo {author} {\bibfnamefont {J.}~\bibnamefont
  {Deseigne}}, \bibinfo {author} {\bibfnamefont {S.}~\bibnamefont
  {L{\'e}onard}}, \bibinfo {author} {\bibfnamefont {O.}~\bibnamefont
  {Dauchot}}, \ and\ \bibinfo {author} {\bibfnamefont {H.}~\bibnamefont
  {Chat{\'e}}},\ }\bibfield  {title} {\enquote {\bibinfo {title} {Vibrated
  polar disks: spontaneous motion, binary collisions, and collective
  dynamics},}\ }\href@noop {} {\bibfield  {journal} {\bibinfo  {journal} {Soft
  Matter}\ }\textbf {\bibinfo {volume} {8}},\ \bibinfo {pages} {5629--5639}
  (\bibinfo {year} {2012})}\BibitemShut {NoStop}%
\bibitem [{\citenamefont {Tsai}\ \emph {et~al.}(2005)\citenamefont {Tsai},
  \citenamefont {Ye}, \citenamefont {Rodriguez}, \citenamefont {Gollub},\ and\
  \citenamefont {Lubensky}}]{tsai2005}%
  \BibitemOpen
  \bibfield  {author} {\bibinfo {author} {\bibfnamefont {J-C}\ \bibnamefont
  {Tsai}}, \bibinfo {author} {\bibfnamefont {Fangfu}\ \bibnamefont {Ye}},
  \bibinfo {author} {\bibfnamefont {Juan}\ \bibnamefont {Rodriguez}}, \bibinfo
  {author} {\bibfnamefont {Jerry~P}\ \bibnamefont {Gollub}}, \ and\ \bibinfo
  {author} {\bibfnamefont {TC}~\bibnamefont {Lubensky}},\ }\bibfield  {title}
  {\enquote {\bibinfo {title} {A chiral granular gas},}\ }\href@noop {}
  {\bibfield  {journal} {\bibinfo  {journal} {Phys.~Rev.~Lett.}\ }\textbf
  {\bibinfo {volume} {94}},\ \bibinfo {pages} {214301} (\bibinfo {year}
  {2005})}\BibitemShut {NoStop}%
\bibitem [{\citenamefont {Crocker}\ and\ \citenamefont
  {Grier}(1996)}]{crocker1996methods}%
  \BibitemOpen
  \bibfield  {author} {\bibinfo {author} {\bibfnamefont {J.~C}\ \bibnamefont
  {Crocker}}\ and\ \bibinfo {author} {\bibfnamefont {D.~G.}\ \bibnamefont
  {Grier}},\ }\bibfield  {title} {\enquote {\bibinfo {title} {Methods of
  digital video microscopy for colloidal studies},}\ }\href@noop {} {\bibfield
  {journal} {\bibinfo  {journal} {J.~Colloid Interface Sci.}\ }\textbf
  {\bibinfo {volume} {179}},\ \bibinfo {pages} {298--310} (\bibinfo {year}
  {1996})}\BibitemShut {NoStop}%
\bibitem [{\citenamefont {Kudrolli}\ \emph {et~al.}(2008)\citenamefont
  {Kudrolli}, \citenamefont {Lumay}, \citenamefont {Volfson},\ and\
  \citenamefont {Tsimring}}]{Kudrolli:2008}%
  \BibitemOpen
  \bibfield  {author} {\bibinfo {author} {\bibfnamefont {A.}~\bibnamefont
  {Kudrolli}}, \bibinfo {author} {\bibfnamefont {G.}~\bibnamefont {Lumay}},
  \bibinfo {author} {\bibfnamefont {D.}~\bibnamefont {Volfson}}, \ and\
  \bibinfo {author} {\bibfnamefont {L.~S.}\ \bibnamefont {Tsimring}},\
  }\bibfield  {title} {\enquote {\bibinfo {title} {Swarming and swirling in
  self-propelled polar granular rods},}\ }\href@noop {} {\bibfield  {journal}
  {\bibinfo  {journal} {Phys. Rev. Lett.}\ }\textbf {\bibinfo {volume} {100}},\
  \bibinfo {pages} {058001} (\bibinfo {year} {2008})}\BibitemShut {NoStop}%
\bibitem [{\citenamefont {Kudrolli}(2010)}]{Kudrolli:2010}%
  \BibitemOpen
  \bibfield  {author} {\bibinfo {author} {\bibfnamefont {A.}~\bibnamefont
  {Kudrolli}},\ }\bibfield  {title} {\enquote {\bibinfo {title} {Concentration
  dependent diffusion of self-propelled rods},}\ }\href@noop {} {\bibfield
  {journal} {\bibinfo  {journal} {Phys. Rev. Lett.}\ }\textbf {\bibinfo
  {volume} {104}},\ \bibinfo {pages} {088001} (\bibinfo {year}
  {2010})}\BibitemShut {NoStop}%
\bibitem [{\citenamefont {Schwager}\ and\ \citenamefont
  {P\"oschel}(2008)}]{Schwager:2008}%
  \BibitemOpen
  \bibfield  {author} {\bibinfo {author} {\bibfnamefont {Thomas}\ \bibnamefont
  {Schwager}}\ and\ \bibinfo {author} {\bibfnamefont {Thorsten}\ \bibnamefont
  {P\"oschel}},\ }\bibfield  {title} {\enquote {\bibinfo {title} {Coefficient
  of restitution for viscoelastic spheres: The effect of delayed recovery},}\
  }\href {\doibase 10.1103/PhysRevE.78.051304} {\bibfield  {journal} {\bibinfo
  {journal} {Phys. Rev. E}\ }\textbf {\bibinfo {volume} {78}},\ \bibinfo
  {pages} {051304} (\bibinfo {year} {2008})}\BibitemShut {NoStop}%
\end{thebibliography}
%

\end{document}